\documentclass[12pt, preprint]{elsarticle}

\usepackage{amsmath,amsfonts,a4,epsfig,graphicx}
\usepackage{amssymb,feynmp,setspace,cancel}
\usepackage{array, tabularx}
\usepackage{float}
\usepackage{color}
\usepackage{verbatim}
\usepackage[colorlinks=false]{hyperref}

\newcommand{\ie}{{\it i.e.}}
\newcommand{\eg}{{\it e.g.}}

\newcommand{\herwig}{{\sc Herwig}}

\newcommand{\madgraph}{{\sc MadGraph}}

\newcommand{\feynrules}{{\sc FeynRules}}
\newcommand{\ufo}{{\sc Ufo}}
\newcommand{\aloha}{{\sc Aloha}}

\newcommand{\helas}{{\sc Helas}}

\newcommand{\cpp}{{\tt C++}}
\newcommand{\fortran}{{\tt Fortran}}
\newcommand{\python}{{\tt Python}}

 \definecolor{lightgrey}{gray}{0.9}

 \def\btab#1\etab{\begin{tabular}{p{50mm}p{70mm}}#1\end{tabular}}
\def\btabx#1\etabx{\begin{tabular}{p{65mm}p{55mm}}#1\end{tabular}}
\def\btaby#1\etaby{\begin{tabular}{p{20mm}p{100mm}}#1\end{tabular}}
\def\btabwide#1\etabwide{\begin{tabular}{p{80mm}p{40mm}}#1\end{tabular}}
\def\btabz#1\etabz{\begin{tabular}{p{40mm}p{25mm}p{25mm}p{30mm}}#1\end{tabular}}
\def\btabp#1\etabp{\begin{tabular}{p{35mm}p{35mm}}#1\end{tabular}}
\def\btabpart#1\etabpart{\begin{tabular}{p{10mm}p{50mm}}#1\end{tabular}}
 \def\bcen{\begin{center}}
 \def\ecen{\end{center}}
 
\def\bgfb#1\egfb{\bcen\fcolorbox{black}{lightgrey}{\parbox{130mm}{\btab#1\etab}}\ecen}
\def\bgfbx#1\egfbx{\bcen\fcolorbox{black}{lightgrey}{\parbox{130mm}{\btabx#1\etabx}}\ecen}
\def\bgfbalign#1\egfbalign{\bcen\fcolorbox{black}{lightgrey}{\parbox{130mm}{\btaby#1\etaby}}\ecen}
\def\bgfbwide#1\egfbwide{\bcen\fcolorbox{black}{lightgrey}{\parbox{130mm}{\btabwide#1\etabwide}}\ecen}
\def\bgfbz#1\egfbz{\bcen\fcolorbox{black}{lightgrey}{\parbox{140mm}{\btabz#1\etabz}}\ecen}
\def\bgfbp#1\egfbp{\bcen\fcolorbox{black}{lightgrey}{\parbox{76mm}{\btabp#1\etabp}}\ecen}
\def\bgfbpart#1\egfbpart{\bcen\fcolorbox{black}{lightgrey}{\parbox{70mm}{\btabpart#1\etabpart}}\ecen}
\newcommand{\etc}[0]{\emph{etc.{}} }
\def\btabmssm#1\etabmssm{\begin{tabular}{p{28mm}p{20mm}}#1\end{tabular}}
\def\bgfbmssm#1\egfbmssm{\bcen\fcolorbox{black}{lightgrey}{\parbox{55mm}{\btabmssm#1\etabmssm}}\ecen}

\def\btabmssmtwo#1\etabmssmtwo{\begin{tabular}{p{32mm}p{44mm}}#1\end{tabular}}
\def\bgfbmssmtwo#1\egfbmssmtwo{\bcen\fcolorbox{black}{lightgrey}{\parbox{82mm}{\btabmssmtwo#1\etabmssmtwo}}\ecen}
\begin{document}
\begin{frontmatter}

\title{
\vspace{-2cm}\begin{flushright}
\small CP3-11-26
\end{flushright}
\vspace{1.5cm}
ALOHA: Automatic Libraries Of Helicity Amplitudes for Feynman diagram computations}

\author[a,b]{\small{Priscila de Aquino}}
\author[c]{\small{William Link}}
\author[b]{\small{Fabio Maltoni}}
\author[b]{\small{Olivier Mattelaer}}
\author[c]{\small{Tim Stelzer}}
\address[a]{Instituut voor Theoretische Fysica, Katholieke Universiteit Leuven, 
\\ Celestijnenlaan 200D, B-3001 Leuven, Belgium}
\address[b]{Center for Cosmology, Particle Physics and Phenomenology (CP3), 
\\ Universit\'e Catholique de Louvain, B-1348 Louvain-la-Neuve, Belgium}
\address[c]{Department of Physics, University of Illinois at Urbana-Champaign, \\ 
1110 West Green Street, Urbana, IL\ \ 61801}

\begin{abstract}
We present an application that automatically writes the \helas~(HELicity Amplitude Subroutines) library corresponding to the Feynman rules of any in quantum field theory Lagrangian. The code is written in \python~ and takes the Universal FeynRules Output (\ufo) as an input. From this input it produces the complete set of routines, wave-functions and amplitudes, that are needed for the computation of Feynman diagrams at leading as well as at higher orders. The representation is language independent and currently it can  output routines in \fortran, \cpp, and \python. A few sample applications implemented in the \madgraph~5 framework are presented.
\end{abstract}

\end{frontmatter}

{\bf PROGRAM SUMMARY}\\

\begin{small}
\noindent
{\em Manuscript Title:} ALOHA: Automatic Libraries Of Helicity Amplitudes for Feynman diagram computations                                       \\ \\
{\em Authors:} Priscila de Aquino, William Link, Fabio Maltoni, Olivier Mattelaer, Tim Stelzer                                                \\\\
{\em Program Title:} ALOHA                                         \\ \\
{\em Journal Reference:}                                      \\ \\
{\em Catalogue identifier:}                                   \\ \\
{\em Licensing provisions: http://www.opensource.org/licenses/UoI-NCSA.php}                                   \\ \\
{\em Programming language:} Python2.6                                   \\ \\
{\em Computer:} 32/64 bit                                              \\ \\
{\em Operating system:} Linux/Mac/Windows                                      \\ \\
{\em RAM:} 512 Mbytes                                              
{\em Classification:} \\
4.4 Feynman diagrams,\\
11.6 Phenomenological and Empirical Models and Theories                                       \\ \\
{\em Nature of problem:}\\
An efficient numerical  evaluation of a squared matrix element can be done with the help of the helicity routines implemented in the HELAS library \cite{1}. 
This static library contains a limited number of helicity functions and is therefore
not always able to provide the needed routine in presence of an arbitrary interaction.
This program provides a way to automatically create the corresponding routines for any given model.
   \\ \\
{\em Solution method:}\\
\aloha~takes the Feynman rules associated to the vertex obtained from the model information (in the \ufo~format \cite{2}), and multiply it by the different wavefunctions or propagators. As a result the analytical expression of the helicity routines is obtained. Subsequently, this expression is automatically written in the requested language (\python, \fortran~or \cpp)\\ \\
{\em Restrictions:}\\
  The allowed fields are currently spin $0$, $1/2$, $1$ and $2$, and the propagators of these particles are canonical.
   \\ \\
{\em Unusual features:}\\
  None
   \\ \\
{\em Running time:}\\
A few seconds for the SM and the MSSM, and up to few minutes for model with spin 2 particles.
   \\ \\

\end{small}

\section{Introduction and motivation}
\label{intr}
We have now entered the era of the Large Hadron Collider (LHC) and measurements based on the Standard Model (SM) predictions have already begun. As data continues to be analysed, deviations from the SM are likely to occur and the challenge will be understanding which Beyond the Standard Model theories (BSM) are responsible for such deviations. Extracting information from the LHC data requires the ability to accurately simulate events based on the theoretical models. While it is possible for researchers to manually create event simulators for the SM and a few popular BSM theories, such an approach would severely restrict the range of models that could be tested at the LHC.

To alleviate the potential bottleneck in simulating BSM theories, extensive efforts are being made to automate the process of going from the Lagrangian of a new theory, to simulating events  that can be then compared with data \cite{Semenov:2010qt,Pukhov:2004ca,Boos:2009un,Brooijmans:2012yi}.   In this paper we present one important piece of that process, \ie, the ability to automatically create helicity amplitude routines necessary for calculating the matrix element associated with any new model.  Such an ability will be directly useful to many groups writing custom generators, however, at present the greatest impact is on the very recently released \madgraph~5 suite~\cite{Alwall:2011uj}.  Prior to this work, \madgraph\  was restricted to generating amplitudes for models that had the same Lorentz structures as the Standard Model. Adding new particles and new interactions, from the MSSM~\cite{Cho:2006sx} to higher spin particles such as gravitons~\cite{Hagiwara:2008jb} or gravitinos~\cite{Hagiwara:2010pi}, required identifying, writing and debugging new helicity amplitudes associated with the model. This severely restricted the range of models that could be studied, as the number of physicists able to implement new models is very limited. For example, technicolor, chiral perturbation theory, and chromomagnetic moments all require extensions to the HELAS library. The ability to dynamically generate the required helicity routines  eliminates the bottleneck, giving all physicists the ability to easily simulate any new model within the \madgraph~5 framework. For instance, all models presently available in the \feynrules\ framework~\cite{Christensen:2008py,Duhr:2011se} can be used via the UFO format~\cite{Degrande:2011ua} in \madgraph~5.

This paper is organized as follows. In Sec.~\ref{sec:helicity} we give a simple example with the aim of explaining the idea of calculating Feynman diagrams routines for vertices and propagators as implemented in \helas. This also allows us to establish a dictionary of terms that will be used in the following sections. In Sec.~\ref{sec:automation} we describe in more detail how the ALOHA program works. In Sec.~\ref{sec:validation} we describe the validation procedure and collect our conclusions in Sec.~\ref{sec:summary}.  

\section{Helicity amplitudes}
\label{sec:helicity}

Helicity amplitudes methods 
\cite{Berends:1981rb, DeCausmaecker:1981bg, Kleiss:1985yh, Gastmans:1990xh, Xu:1986xb, Gunion:1985vca, Hagiwara:1985yu} are a convenient and effective way to evaluate the squared matrix element of any process.  As the name implies, helicity amplitude methods work at the amplitude level, in contrast to the trace techniques based on completeness relations which work on squared amplitudes. This has two important advantages. First, the complexity in the calculation grows linearly with the number of diagrams instead of quadratically. Second, it is possible to ``factorize"  diagrams, such that if a particular substructure shows up in several diagrams, it only needs to be calculated once, significantly increasing the speed of the calculation (see for example \cite{Alwall:2011uj}).   

The \helas~\cite{Murayama:1992gi, Hagiwara:2008zr} library has been a particularly successful implementation of an helicity amplitude method for tree-level processes.  Amplitudes are generated by initializing a set of external wavefunctions using their helicity and momenta. These wavefunctions are combined based on the particle interactions in the Lagrangian to determine the wavefunctions of the internal lines (propagators).  Once all of the wavefunctions are determined, they are combined to calculate the complex number corresponding to the amplitude for the diagram. These amplitudes can then be added and squared to give the required result. The set of routines needed to calculate amplitudes for the Standard Model at tree-level was released in the original \helas~package \cite{Murayama:1992gi} and successfully employed for many phenomenological studies thereafter. 

\begin{figure}
\center
\includegraphics[scale=0.25]{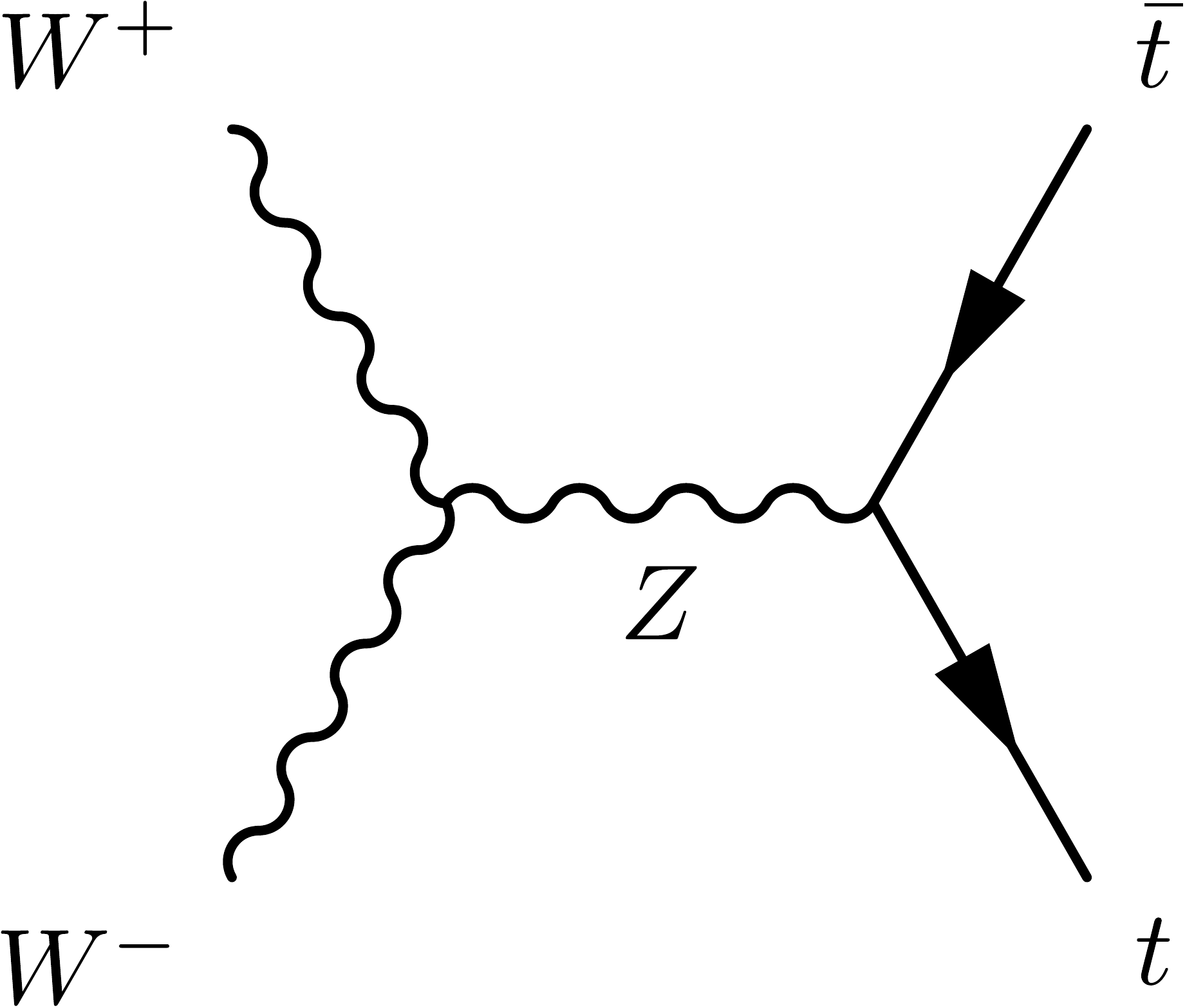}
\vspace{0.3 in}
\caption{ \label{fig:WWtbart} $W^+W^-\to t\bar t$ via $Z$ exchange }
\end{figure}

In addition to having  a complete and highly optimized set of routines, the structure of the \helas~calls makes it particularly easy to write (and read) the code for any Feynman diagram: one line of code for each line of the associated Feynman diagram, and a final call that returns the amplitude. For example, the diagram for $W^+W^-\to t \bar t$ via a $Z$ boson, shown in Fig.~\ref{fig:WWtbart}, is written in six lines.
First, there are calls to functions associated with the external particles\footnote{The notation follows the \helas~convention  \cite{Murayama:1992gi}.}:
\begin{verbatim}
   call VXXXXX(P_WM, W_MASS, WM_HEL, -1, WM)
   call VXXXXX(P_WP, W_MASS, WP_HEL, -1, WP)
   call OXXXXX(P_T, T_MASS, T_HEL, +1, FO)
   call IXXXXX(P_TBn T_MASS, TB_HEL, -1, FI). 
\end{verbatim} 
From the four resulting wavefunctions (\texttt{WM, WP, FO, FI}) one can combine either the two fermions or the two bosons in order to build the propagator for the $Z$ boson. Choosing to connect the fermions, the code reads:
\begin{verbatim}
   call J3XXXX(FI, FO, GAU, GZU, Z_MASS, Z_WIDTH, J3).
\end{verbatim}
Finally, the $Z$ boson can be combined with the external bosons to evaluate the amplitude:
\begin{verbatim}
   call VVVXXX( WP, WM, J3, GW, AMP).
\end{verbatim}

The example above demonstrates that \helas~consists of three categories of routines:  
\begin{description}
\item{\emph {External particle routines:}} Each external leg of a Feynman diagram corresponds to a wavefunction computed for a given momentum and a given helicity. For example, the first four routines used in the $W^+W^-\to t \bar t$ calculation are of this class. The only input expected for this category of routines is the type of the particle (\ie~vector, scalar, tensor or incoming/outgoing fermion), is independent of the interactions in the model, and depends only on the Lorentz representation of the particles present in the theory.

\item{\emph{Wavefunction/Off-shell routines:}} When one leg of a vertex has no associated wavefunction, a routine is used to generate the internal particle (\ie, propagator) wavefunction. This is exemplified by the $Z$ boson call above. These routines return both the internal wavefunction and the momentum of the propagator. 

\item{\emph{Amplitude/On-shell routines:}} When the wavefunctions for all the legs at a vertex are known, they can be used to compute the matrix-element of the diagram by the associated amplitude routines. 
\end{description}

To simulate new models, relatively few external particles are required, as they depend only on the Lorentz representation of the particle. However, the wavefunction and amplitude routines depend explicitly on the interactions of the theory, making it impractical to anticipate and create a comprehensive library of routines for all interesting BSM theories. 

\section{Automation of the method}
\label{sec:automation}

Originally, the \helas~package contained external particle routines only for classes of particles present in the Standard Model (particles with spin 0, 1 and 1/2). Recently the library has been expanded to include spin-2 particles \cite{Hagiwara:2008zr, deAquino:2011ix}, spin-3/2 particles \cite{Hagiwara:2010pi, Mawatari:2011jy}, as well as the Higgs effective theory \cite{Alwall:2007st}.  This set of external particle routines is sufficient for calculating most renormalizable BSM theories of interest.  

The challenge that remains is writing the required wavefunction and amplitude routines required for any BSM theory, renormalizable or not. Although the rules for writing such a routine are known, it can be very time consuming to verify that all of the sign conventions and normalizations have been properly implemented.  This provides a severe limitation to \helas-based event generators.  To fill this need, we develop \aloha: the {\tt A}utomatic {\tt L}anguage-independent {\tt O}utput of {\tt H}elicity {\tt A}mplitudes, which is capable of automatically generating \helas~routines in \fortran, \cpp~and \python. 

\subsection{Computation method}

\aloha~is written in \python, the same language used by the \ufo~code \cite{Degrande:2011ua} to ensure maximum flexibility and compatibility.  Moreover, the input model file used by \aloha~is the one provided by the \ufo.

In order to generate the necessary \helas~routines, \aloha~needs to perform specific analytic computations. A package to deal with these mandatory computations has been developed and its details can be found in  \ref{app:analytical}. 

We begin by describing the relevant information existing in the \ufo~code, which is the starting point of \aloha. All information concerning the model, \ie~particles, interactions and parameters, is obtained from the \ufo. \aloha~connects particles, creating Lorentz structures to output the wavefunctions and amplitudes using the syntax defined by the original \helas~package.  

To illustrate how \aloha~works, consider the $e^ + e^- \gamma$ vertex produced by \ufo:

\begin{verbatim}
FFV = Lorentz(name = 'FFV',
              spins = [ 2, 2, 3 ],
              structure = 'Gamma(3,2,1)')
\end{verbatim}
The {\tt spins} attribute contains the list of spins of each particle (present in the vertex) in the {\tt 2*S+1} scheme\footnote{In the case of a fermion, the first particle is considered as an incoming fermion and the second as an outgoing fermion. For interactions containing more than two fermions, the convention is that the vertex should be defined for a given fermion flow. The outgoing fermion follows directly its associated incoming fermion.}. The {\tt structure} attribute collects the analytical expression for the Lorentz structure in the vertex.  The arguments of the Lorentz object refer to the particle associated with each index of the object.  The list of available objects and the \python~convention for different indices are explained in Table~\ref{tab:lorentz_structure}.
\begin{table}
\bgfbwide
\multicolumn{2}{c}{\textbf{Table~\ref{tab:lorentz_structure}: Elementary Lorentz structures}}\\
\\
Charge conjugation matrix: $C_{a b}$& {\tt C(a,b)}\\ 
Epsilon matrix: $\epsilon^{\mu\nu\rho\sigma}$ & {\tt Epsilon($\mu$,$\nu$,$\rho$,$\sigma$)}\\
Gamma matrices: $(\gamma^{\mu})_{ab}$ & {\tt Gamma($\mu$, a, b)}\\
Gamma5 matrix:  $(\gamma^5)_{ab}$ & {\tt Gamma5(a,b)} \\
(Spinorial) Kronecker delta: $\delta_{a b}$ &  {\tt Identity(a,b)} \\
Minkowski metric: $\eta_{\mu \nu}$ & {\tt  Metric($\mu$,$\nu$) }\\
Momenta of particle $N$: $p^{\mu}_N$& {\tt  P($\mu$,N) }\\
Right chiral projector:  $\left(\frac{1+\gamma5}{2}\right)_{a b}$ & {\tt  ProjP(a,b) }\\
Left chiral projector $\left(\frac{1-\gamma5}{2}\right)_{a b}$ &  {\tt ProjM(a,b) }\\
Sigma matrices: $(\sigma^{\mu\nu})_{a b}$ &  {\tt Sigma($\mu$,$\nu$,a,b)}\\
Scalar   wavefunction: $ \phi_N$& Scalar(N)\\
Spinor wavefunction: $ \psi^a_N$ & Spinor(a,N)\\
Vector  wavefunction:  $ \epsilon_\mu^N$ & Vector($\mu$,N)\\
Spin2   wavefunction: $T^{\mu,\nu}_N$&  Spin2($\mu$,$\nu$,N)\\
\egfbwide

\caption{\label{tab:lorentz_structure} The greek indices stands for lorentz indices while the lower case latin indices stands for spin indices. The indices $N$ stands for the label of the particle.}
\end{table}
The \ufo~convention requires:
\begin{itemize}
\item positive indices to be linked to the particle numbers;
\item all Lorentz indices to be listed before spin indices;
\item all repeated indices to be summed with the appropriate matrix;
\item all momenta to be incoming. Note this is different for the \helas/\aloha~convention.
\end{itemize}

From a given \ufo~model structure, \aloha~creates the analytical expression linked to the associated \helas~routine by contracting the expression with a set of wavefunctions.  To illustrate how it works we use the structure shown in the {\tt FFV} python object introduced above: {\tt Gamma(3,2,1)}. 

The associated amplitude routine, which was previously named {\tt IOV} by the \helas~convention, is now:
\begin{verbatim} 
Spinor(2) * Gamma(3,2,1) * Spinor(1) * Vector(3)
\end{verbatim}
which correspond to the analytical expression: 
\begin{equation}
\label{ffv0}
{\tt FFV\_0}(\psi_1,\psi_2,\epsilon_{3\,\mu}, g) =g * \bar u_2 \gamma^\mu v_1 \epsilon_\mu.
\end{equation}
The analytical module of \aloha~expands Eq.~(\ref{ffv0}) component by component. The result is written to a \cpp, \python~or \fortran~file, language chosen by the user. For instance, the Fortran output would be:
\begin{verbatim}
C     This File is Automatically generated by ALOHA                             
C     The process calculated in this file is:                                   
C     Gamma(3,2,1)                                                              
C                                                                               
      SUBROUTINE FFV1_0(F1,F2,V3,COUP,VERTEX)
      IMPLICIT NONE
      DOUBLE COMPLEX F1(*)
      DOUBLE COMPLEX F2(*)
      DOUBLE COMPLEX V3(*)
      DOUBLE COMPLEX COUP
      DOUBLE COMPLEX VERTEX

      VERTEX = COUP*( (F2(1)*( (F1(3)*( (0, -1)*V3(1)+(0, 1)*V3(4)))
     $ +(F1(4)*( (0, 1)*V3(2)+V3(3)))))+( (F2(2)*( (F1(3)*( (0, 1)
     $ *V3(2)-V3(3)))+(F1(4)*( (0, -1)*V3(1)+(0, -1)*V3(4)))))
     $ +( (F2(3)*( (F1(1)*( (0, -1)*V3(1)+(0, -1)*V3(4)))+(F1(2)
     $ *( (0, -1)*V3(2)-V3(3)))))+(F2(4)*( (F1(1)*( (0, -1)*V3(2)
     $ +V3(3)))+(F1(2)*( (0, -1)*V3(1)+(0, 1)*V3(4))))))))
      END
\end{verbatim} 

\aloha's naming convention for generated routines is denoted by \texttt{NAME\_X}, where \texttt{NAME} is the name provided by \ufo~and {\tt X} is either {\tt 0} for an amplitude routine or the number corresponding to the off-shell particle for a wavefunction routine. 
The convention for the arguments is detailed in \ref{app:convention}.  

To produce the necessary off-shell routines, \aloha~loops over the list of particles and replaces the on-shell wavefunction with a propagator.  For particle 1, the incoming spinor is replaced by a propagator, computing\footnote{The propagator object is defined internally to \aloha.}: 
\begin{verbatim} 
Spinor(2) * Gamma(3,2,1) * SpinorPropagator(1) * Vector(3)
\end{verbatim}
which is analogous to:
\begin{equation}
{\tt FFV\_1}(\psi_2,\epsilon_{3\,\mu}, g, m_1, \Gamma_1) = g * \bar{\psi_2}  \, \gamma^{\mu}\cdot i\frac{\cancel{p} + m_1}{p^2-m_1^2+im_1\Gamma_1} \, \epsilon_{3\mu}.
\end{equation} 
{\tt FFV\_1} is the wavefunction of the first spinor, and would be {\tt FOV} in the original \helas~language. The momentum for the propagator is computed by using  conservation of momentum at the vertex. 

Similarly, for the outgoing spinor, we have: 
\begin{equation}
{\tt FFV\_2}(\psi_1,\epsilon_{3\,\mu},g,m_2, \Gamma_2) = i \, \frac{\cancel{p} + m_2}{p^2-m_2^2+im_2\Gamma_2}\cdot i \, \gamma^{\mu} \, \psi_2 \, \epsilon_{3\mu}.
\end{equation}
The syntax used by \aloha~reads:
\begin{verbatim}
SpinorPropagator(2) * Gamma(3,2,1)  * Spinor(1) * Vector(3)
\end{verbatim}
which can be related to the {\tt FIV} routine in the original \helas~language. 

Finally, \aloha~also creates the routine associated to the off-shell vector.   This routine  returns a vector wavefunction from a pair of incoming and outgoing fermion wavefunctions, and it is written by \aloha~as:
\begin{verbatim}
Spinor(2) * Gamma(3,2,1) Spinor(1) * VectorPropagator(3)
\end{verbatim}
analogous to the analytic expression {\tt FFV\_3}, which would be {\tt JIO} in the original \helas~language:
\begin{equation} 
{\tt FFV\_3}(\psi_1,\psi_2, g, m_3, \Gamma_3) = \bar{\psi_2} \, i \, \gamma_{\mu} \, \psi_1 \cdot-i \, \frac{\eta^{\mu\nu}-\frac{p^\mu p^\nu}{m_3^2}}{p^2-m_3^2+im_3\Gamma_3}.
\end{equation}

Following the original \helas~convention, \aloha~uses the unitary gauge for massive particles and the light cone gauge for massless particles.  
Notice  that \aloha~is not designed to create external particles routines, \eg, {\tt IXXXXX} and {\tt OXXXXX} for incoming and outgoing fermions, {\tt VXXXXX} for vectors, {\tt SXXXXX} for scalars and  {\tt TXXXXX} for spin-2 particles. There is a relatively small number of routines needed, and it is therefore reasonable to use a static library\footnote{These external particle routines are available in \python, \cpp~and \fortran~languages in the \aloha~package.}.

In order to cover the full range of routines required for any type of interaction (and model), \aloha~still needs two additional types of special routines:
\begin{itemize}

\item {\bf Conjugate Routines:}~The presence of majorana fermions and of fermion flow violation, requires for some processes the presence of the conjugated version of the routines \cite{Denner:1992vza} not present in the Standard Model. In the case of the previous example, the Lorentz expression would have to be replaced by:
\begin{verbatim}
C(-2,2) * Spinor(2) * Gamma(3,2,1) *
                              Spinor(1) * Vector(3) * C(3,-3) 
\end{verbatim}
In order to distinguish these conjugate routines from the standard ones, \aloha~adds to their name the letter {\tt C}  followed by the number of the fermion pair that is conjugated. In this case the name of the conjugate routine would be {\tt FFVC1\_X} with {\tt X} being either {\tt 0, 1, 2} or {\tt 3} depending on the number corresponding to the amplitude or off-shell particle. Note that {\tt \_1} ({\tt \_2}) returns the wavefunctions associated to the incoming (outcoming) fermion.

\item {\bf Multiple Couplings:} In the \ufo~scheme each Lorentz structure is associated to a single coupling. In practice, this means that some interactions are linked to more than one Lorentz structure (for example, this is the case for the $Z$ boson in the Standard Model). However, it is convenient for the readability and the compactness of the code to have one single routine per interaction. Therefore, \aloha~defines a series of wrapper functions which call different routines associated by their couplings. The name of these wrapper routines are the concatenation of the name of the associated routines combined by an underscore. If all of the associated routines start with the same prefix (with letter S, F, V, T) then this prefix is only kept the first time. For example, the wrapper function of  {\tt FFV1\_X} and  {\tt FFV2\_X} will be called  {\tt FFV1\_2\_X}.
\end{itemize}

In some cases, this method provides redundancy. Consider a four-Higgs-interaction. Due to the symmetry of the Lorentz structure, the four wavefunction routines will each have the same content. This type of symmetry is detected automatically by \aloha~and the computation is done only once. However, in order to simplify the work of the matrix element generator, all four routines are explicitly defined and have their own names. The first one is computed and the remaining three are written as an alias of the first one.

\subsection{Resulting output}
 
The major advantage of \aloha~comes from using it in combination with a Monte Carlo (MC) event generator. To this aim, \aloha~offers two options. 
The first is to use \aloha~as a standalone program.  The user can call \aloha~from the shell:
\begin{verbatim}
./bin/aloha UFO_path [ -f  Fortran | Python | CPP] [-o output_dir]
\end{verbatim}
where {\tt UFO\textunderscore path} is the path to the \ufo~model directory, and the two possible arguments are 
\begin{itemize}
\item the language in which the routine should be written (default is \fortran). 
\item the path for the output directory (which is by default {\tt UFO\_path/LANGUAGE}, where {\tt LANGUAGE} can be either \fortran, \python~or \cpp).
\end{itemize}

The second possibility is to internally link the \aloha~code to a MC event generator. This option is only possible for a generator written in \python, and  
is currently used by \madgraph~version 5.  For each process \madgraph~requests the required set of helicity routines, which is produced by \aloha. This first implementation of \aloha~inside a matrix element generator played an essential role in fully validating the package.

\section{Tests and validation}
\label{sec:validation}

The validity and precision of the original \helas~routines has been known for many years. A natural choice for testing \aloha~routines is therefore to compare them with the original \helas~routines.
Since the format of the \aloha~is different from that of \helas, a line-by-line comparison would not work as a general validation test. Instead, validation is provided by the calculation of amplitudes of a particular process using the two sets of routines and comparing results. Below is a full description for the validation tests, as well as an example of the power and utility provided by such tests.

\subsection{The validation process}

\madgraph~version 4 uses \helas~routines to compute values for the square of the matrix element for a specific phase-space point of a known process. New routines can be validated by computing these values using both the original \helas~and the new \aloha~routines.

The idea is to use two different versions of \madgraph~to compute these values. 
We employ \madgraph~version 4.4.44, which uses the original \helas~routines, and compare the results with those from  \madgraph~version 5.1.1.0, which uses the new \aloha~routines. 

The comparison is made by specifying a point of phase space (randomly generated by {\tt{RAMBO}} \cite{Kleiss:1985gy}) and a center-of-mass energy, after fixing properly the set of external parameters. The values for each matrix element squared are computed, and the difference is given by $\Delta D$:
\begin{equation}
\Delta D= 2 \times \frac{|\mathcal{M} (\tt{ALOHA})|^ 2 - |\mathcal{M}(\tt{HELAS})|^ 2}{|\mathcal{M}(\tt{ALOHA})|^ 2 + |\mathcal{M}(\tt{HELAS})|^ 2}.
\label{DeltaD}
\end{equation}
Due to effects of numerical precision, one can not expect exact agreement between the two codes. Instead, the test is considered to pass if values of the squared matrix element computed by \aloha~and \helas~agree to at least five orders of magnitude, \ie, $\Delta D \le 1 \times 10^{-5}$. 

These tests are carried out for different processes based on the Standard Model (SM), the Minimal Supersymmetric Model (MSSM) and the Randall-Sundrum type I (RS-I) model, the later including spin-2 particles. Additional tests are made by checking gauge and Lorentz invariance of each process. We observe a great agreement for all processes, as illustrated in Tables \ref{tab:SM2to2results} - \ref{tab:RS2to3results} and explained below.

\subsection{Standard Model processes}
\label{Session:sm}

For the Standard Model, we test a very large sample of processes.
Namely around 800 $2 \rightarrow 2$, 600 $2 \rightarrow 3$ and 50 $2 \rightarrow 4$ processes.
These are chosen to ensure that all helicity routines have been covered.

Some examples of these results can be found in Tables \ref{tab:SM2to2results} and \ref{tab:SM2to3results}. In both tables, the first column shows the process chosen for the comparison. The second column shows the difference $\Delta D$ given by Eq.~(\ref{DeltaD}) for a specific point of phase space. 

\begin{table}
\begin{minipage}[t]{0.5\linewidth}\centering
\bgfbp
\multicolumn{2}{c}{\textbf{Table~\ref{tab:SM2to2results}: SM $2\rightarrow 2$ results}}\\
\\
{\bf Process} & ${\mathbf  \Delta D}$ \\ \\
$e^+ \, e^- \rightarrow \gamma \, \gamma$ & 1.512$\times 10^ {-7}$\\ 
$e^+ \, e^- \rightarrow W^+ \, W^-$ & 3.376$\times 10^ {-8}$\\ 
$e^- \, \bar \nu_e \rightarrow \gamma \, W^-$ & 4.541$\times 10^ {-8}$\\ 
$\gamma \, e^- \rightarrow \nu_e \, W^-$ & 4.545$\times 10^ {-8}$\\ 
$b~ \, e^- \rightarrow b~ \, e^-$ & 7.310$\times 10^ {-8}$\\ 
$e^- \, W^- \rightarrow e^- \, W^-$ & 5.557$\times 10^ {-8}$\\ 
$e^+ \, e^+ \rightarrow e^+ \, e^+$ & 9.974$\times 10^ {-8}$\\ 
$e^+ \, \bar \nu_e \rightarrow e^+ \, \bar \nu_e$ & 3.911$\times 10^ {-8}$\\ 
$e^+ \, t \rightarrow e^+ \, t$ & 1.317$\times 10^ {-7}$\\ 
$e^+ \, W^+ \rightarrow e^+ \, W^+$ & 5.557$\times 10^ {-8}$\\ 
$e^+ \, Z \rightarrow \gamma \, e^+$ & 5.458$\times 10^ {-8}$\\ 
$\nu_e \, \bar \nu_e \rightarrow e^+ \, e^- $ & 7.164$\times 10^ {-8}$\\ 
$\nu_e \, \bar \nu_e \rightarrow W^+ \, W^-$ & 7.477$\times 10^ {-8}$\\ 
$t~ \, \nu_e \rightarrow t~ \, \nu_e$ & 1.068$\times 10^ {-7}$\\ 
$\nu_e \, W^- \rightarrow \gamma \, e^- $ & 4.545$\times 10^ {-8}$\\ 
$\nu_e \, Z \rightarrow \nu_e \, Z$ & $<$ 1.0$\times 10^ {-10}$\\ 
$t~ \, \bar \nu_e \rightarrow t~ \, \bar \nu_e$ & 8.411$\times 10^ {-8}$\\ 
$\bar \nu_e \, W^+ \rightarrow e^+ \, Z$ & 5.586$\times 10^ {-8}$\\ 
$\bar \nu_e \, Z \rightarrow \bar \nu_e \, Z$ & $<$ 1.0$\times 10^ {-10}$\\ 
$\gamma \, g \rightarrow t \, t~$ & 7.528$\times 10^ {-8}$\\ 
$g \, t \rightarrow t \, Z$ & 3.173$\times 10^ {-8}$\\ 
$b \, g \rightarrow b \, g$ &$<$ 1.0$\times 10^ {-10}$ \\ 
$b~ \, g \rightarrow \gamma \, b~$ & 7.665$\times 10^ {-8}$\\ 
$\gamma \, \gamma \rightarrow e^+ \, e^- $ & 1.512$\times 10^ {-7}$\\ 
$\gamma \, t \rightarrow g \, t$ & 7.611$\times 10^ {-8}$\\ 
$\gamma \, t~ \rightarrow \gamma \, t~$ & 1.524$\times 10^ {-7}$\\ 
$\gamma \, b \rightarrow b \, Z$ & 4.564$\times 10^ {-8}$\\ 
$\gamma \, Z \rightarrow b \, b~$ & 4.554$\times 10^ {-8}$\\ 
$t \, t~ \rightarrow \nu_e \, \bar \nu_e$ & 5.602$\times 10^ {-8}$\\ 
$t \, t~ \rightarrow \gamma \, \gamma$ & 1.530$\times 10^ {-7}$\\ 
$t \, t~ \rightarrow W^+ \, W^-$ & 1.833$\times 10^ {-8}$\\ 
$b \, Z \rightarrow \gamma \, b$ & 4.564$\times 10^ {-8}$\\ 
$b~ \, W^- \rightarrow b~ \, W^-$ & 6.291$\times 10^ {-9}$\\ 
$W^+ \, W^- \rightarrow \gamma \, Z$ & 1.876$\times 10^ {-8}$\\ 
$W^+ \, W^- \rightarrow Z \, Z$ & 1.666$\times 10^ {-7}$\\ 
$Z \, Z \rightarrow e^+ \, e^- $ & 1.162$\times 10^ {-7}$\\ 
\egfbp
\textcolor{white}{\caption{\label{tab:SM2to2results}}}
\end{minipage}
\hspace{0.5cm}
\begin{minipage}[t]{0.5\linewidth}
\centering
\bgfbp
\multicolumn{2}{c}{\textbf{Table~\ref{tab:SM2to3results}: SM $2 \rightarrow 3$ results}}\\
\\
{\bf Process} & ${\mathbf  \Delta D}$ \\ \\
$e^+ \, e^- \rightarrow H \, H \, Z$ & 2.029$\times 10^ {-8}$\\ 
$e^- \, g \rightarrow e^- \, u \, \bar u$ & 4.934$\times 10^ {-8}$\\ 
$d \, e^- \rightarrow d \, e^-  \, H$ & 1.712$\times 10^ {-7}$\\ 
$e^- \, t~ \rightarrow t~ \, \nu_e \, W^-$ & 6.658$\times 10^ {-8}$\\ 
$e^- \, W^- \rightarrow e^- \, W^- \, Z$ & 9.767$\times 10^ {-8}$\\ 
$e^+ \, \nu_e \rightarrow u \, \bar u \, W^+$ & 8.593$\times 10^ {-8}$\\ 
$\gamma \, e^+ \rightarrow \bar \nu_e \, W^+ \, Z$ & 7.611$\times 10^ {-9}$\\
$\nu_e \, Z \rightarrow \gamma \, e^-  \, W^+$ & 2.692$\times 10^ {-8}$\\ 
$u \, \bar \nu_e \rightarrow \gamma \, u \, \bar \nu_e$ & 1.179$\times 10^ {-7}$\\ 
$t \, \bar \nu_e \rightarrow e^+ \, t \, W^-$ & 8.095$\times 10^ {-8}$\\ 
$\bar \nu_e \, W^+ \rightarrow b \, b~ \, e^+$ & 8.430$\times 10^ {-8}$\\ 
$g \, g \rightarrow \gamma \, d \, \bar d$ & 7.526$\times 10^ {-8}$\\ 
$b~ \, g \rightarrow \gamma \, b~ \, g$ & 7.415$\times 10^ {-8}$\\ 
$\gamma \, u \rightarrow e^+ \, e^-  \, u$ & 1.662$\times 10^ {-7}$\\ 
$\gamma \, d \rightarrow d \, e^+ \, e^- $ & 6.045$\times 10^ {-8}$\\ 
$\gamma \, H \rightarrow e^+ \, e^-  \, Z$ & 5.547$\times 10^ {-8}$\\ 
$\gamma \, t \rightarrow t \, t \, t~$ & 7.878$\times 10^ {-8}$\\ 
$u \, \bar u \rightarrow e^+ \, e^-  \, H$ & 1.487$\times 10^ {-7}$\\ 
$u \, \bar u \rightarrow u \, \bar u \, Z$ & 9.593$\times 10^ {-8}$\\ 
$H \, u \rightarrow H \, u \, Z$ & 4.478$\times 10^ {-8}$\\ 
$u \, W^+ \rightarrow e^+ \, u \, \nu_e$ & 8.576$\times 10^ {-8}$\\ 
$\bar u \, \bar u \rightarrow H \, \bar u \, \bar u$ & 1.145$\times 10^ {-7}$\\ 
$t \, \bar u \rightarrow H \, t \, \bar u$ & $<$ 1.0$\times 10^ {-10}$\\ 
$\bar u \, Z \rightarrow e^+ \, e^-  \, \bar u$ & 3.816$\times 10^ {-8}$\\ 
$d \, \bar d \rightarrow H \, \nu_e \, \bar \nu_e$ & 2.566$\times 10^ {-8}$\\ 
$H \, Z \rightarrow g \, t \, t~$ & 2.457$\times 10^ {-8}$\\ 
$t \, t~ \rightarrow \gamma \, \nu_e \, \bar \nu_e$ & 2.065$\times 10^ {-8}$\\ 
$t \, t~ \rightarrow u \, \bar u \, Z$ & 3.962$\times 10^ {-8}$\\ 
$t \, W^- \rightarrow e^- \, t \, \bar \nu_e$ & 4.744$\times 10^ {-8}$\\ 
$b \, t~ \rightarrow \gamma \, b \, t~$ & 7.409$\times 10^ {-8}$\\ 
$t~ \, Z \rightarrow d \, \bar d \, t~$ & 2.224$\times 10^ {-8}$\\ 
$b \, b~ \rightarrow g \, H \, H$ & $<$ 1.0$\times 10^ {-10}$\\ 
$b \, b~ \rightarrow Z \, Z \, Z$ & 9.055$\times 10^ {-8}$\\ 
$b~ \, b~ \rightarrow \gamma \, b~ \, b~$ & 7.375$\times 10^ {-8}$\\ 
$b~ \, Z \rightarrow b~ \, Z \, Z$ & 8.196$\times 10^ {-8}$\\ 
$W^+ \, W^- \rightarrow d \, \bar d \, H$ & 9.519$\times 10^ {-8}$\\ 
\egfbp
\end{minipage}
\textcolor{white}{\caption{\label{tab:SM2to3results}}}
\end{table}

\subsection{MSSM processes}
One common approach to extend the Standard Model is the 4-dimensional Supersymmetric theory, based on the popular field description delineated by Wess and Zumino \cite{Wess:1974tw}. Within this category of  Beyond Standard Model theories is the Minimal Supersymmetric Standard Model, which relates bosons to fermions by unifying symmetries in the most straightforward way \cite{Haber:1984rc, Nilles:1983ge, Rosiek:1989rs, Martin:1997ns}. 
It addresses the Hierarchy Problem by adding new states whose contribution
effectively remove the quadratic divergences to the Higgs mass. It also provides a natural candidate for dark matter if the presence of an additional symmetry, R-parity, is assumed.
On account of being a popular extension to the Standard Model, the MSSM has been extensively searched for at Hadron Colliders and several model implementations are available. We can therefore use these implementations to test \aloha~within a theory which uses the full set of implemented subroutines. Testing \aloha~for the MSSM is similar to testing it for the SM, since the Lorentz structures are for the most identical. However, the presence of majorana fermions and of fermion flow violations, require  the presence conjugated versions of the routines \cite{Denner:1992vza} not present in the Standard Model.
The same validation tests used in Sec.~\ref{Session:sm} for the Standard Model are therefore performed for the MSSM. Approximately 1000 $2 \rightarrow 2$ processes and 7000 $2 \rightarrow 3$ processes have been tested. The results are conclusive and 100\% agreement is obtained. A sample of processes checked for the MSSM is shown in Tables \ref{tab:MSSM2to2results} and \ref{tab:MSSM2to3results}. 

\begin{table}
\begin{minipage}[t]{0.5\linewidth}\centering
\bgfbp
\multicolumn{2}{c}{\textbf{Table~\ref{tab:MSSM2to2results}: MSSM $2 \rightarrow 2$ results}}\\
\\
{\bf Process} & ${\mathbf \Delta D}$ \\ \\
$e^- \, W^+ \rightarrow e^- \, W^+$ & 9.451$\times 10^ {-10}$\\ 
$e_R^+ \, W^+ \rightarrow e_R^+ \, W^+$ & 5.544$\times 10^ {-8}$\\ 
$H_2 \, W^+ \rightarrow H_3 \, W^+$ & 1.108$\times 10^ {-8}$\\ 
$H^- \, W^+ \rightarrow H_2 \, H_3$ & 9.787$\times 10^ {-7}$\\ 
$e_L^+ \, W^- \rightarrow e_L^+ \, H^-$ & 5.285$\times 10^ {-9}$\\ 
$W^- \, x_1^+ \rightarrow H^+ \, x_1^-$ & 2.493$\times 10^ {-6}$\\ 
$H^+ \, W^- \rightarrow H^+ \, W^-$ & 2.055$\times 10^ {-7}$\\ 
$\nu_e \, \bar \nu_e \rightarrow H^+ \, H^-$ & 6.214$\times 10^ {-9}$\\ 
$s\bar \nu_e \, \nu_e \rightarrow e^+ \, e_L^-$ & 1.038$\times 10^ {-7}$\\ 
$H_3 \, \nu_e \rightarrow e_L^- \, x_1^+$ & 1.075$\times 10^ {-7}$\\ 
$e_L^+ \, \bar \nu_e \rightarrow e_L^+ \, \bar \nu_e$ & 5.987$\times 10^ {-8}$\\ 
$H_1 \, \bar \nu_e \rightarrow e^+ \, H^-$ & 8.885$\times 10^ {-7}$\\ 
$e^+ \, e^- \rightarrow s\nu_e \, s\bar \nu_e$ & 8.665$\times 10^ {-8}$\\ 
$e^- \, s\bar \nu_e \rightarrow H_3 \, x_1^-$ & 1.613$\times 10^ {-7}$\\ 
$e^- \, H_1 \rightarrow e^- \, H_3$ & 8.890$\times 10^ {-7}$\\ 
$e^+ \, e_L^- \rightarrow s\nu_e \, \bar \nu_e$ & 1.792$\times 10^ {-9}$\\ 
$e^+ \, e_R^- \rightarrow H^- \, x_1^+$ & 6.679$\times 10^ {-9}$\\ 
$e^+ \, H^+ \rightarrow e_L^+ \, x_1^+$ & 6.977$\times 10^ {-9}$\\ 
$H_1 \, \mu^+ \rightarrow H_3 \, \mu^+$ & 8.890$\times 10^ {-7}$\\ 
$e_L^+ \, e_L^- \rightarrow \mu^+ \, \mu^-$ & 1.984$\times 10^ {-9}$\\ 
$e_L^- \, x_1^+ \rightarrow H_2 \, \nu_e$ & 3.028$\times 10^ {-8}$\\ 
$e_L^- \, H^+ \rightarrow H_1 \, s\nu_e$ & 8.264$\times 10^ {-7}$\\ 
$e_L^+ \, x_1^- \rightarrow H_3 \, \bar \nu_e$ & 1.075$\times 10^ {-7}$\\ 
$e_L^+ \, H^- \rightarrow H_2 \, s\bar \nu_e$ & 1.871$\times 10^ {-9}$\\ 
$s\nu_e \, x_1^- \rightarrow \nu_e \, W^-$ & 3.714$\times 10^ {-8}$\\ 
$H^- \, s\nu_e \rightarrow \nu_e \, x_1^-$ & 1.039$\times 10^ {-8}$\\ 
$H_2 \, s\bar \nu_e \rightarrow e_L^+ \, H^-$ & 1.871$\times 10^ {-9}$\\ 
$e_R^+ \, e_R^- \rightarrow s\nu_e \, s\bar \nu_e$ & 4.856$\times 10^ {-9}$\\ 
$e_R^+ \, H^- \rightarrow e_R^+ \, W^-$ & 7.992$\times 10^ {-9}$\\ 
$x_1^+ \, x_1^+ \rightarrow x_1^+ \, x_1^+$ & 2.863$\times 10^ {-8}$\\ 
$H_2 \, x_1^+ \rightarrow H_3 \, x_1^+$ & 2.019$\times 10^ {-8}$\\ 
$H_2 \, x_1^- \rightarrow e^- \, s\bar \nu_e$ & 1.527$\times 10^ {-7}$\\ 
$H_1 \, H_1 \rightarrow H_1 \, H_1$ & 1.139$\times 10^ {-8}$\\ 
$H_1 \, H_3 \rightarrow H_2 \, H_3$ & 8.991$\times 10^ {-7}$\\ 
$H_2 \, H_2 \rightarrow H_3 \, H_3$ & 1.021$\times 10^ {-8}$\\ 
\egfbp
\textcolor{white}{\caption{\label{tab:MSSM2to2results}}}
 \end{minipage}
\hspace{0.5cm}
\begin{minipage}[t]{0.5\linewidth}
\centering
\bgfbp
\multicolumn{2}{c}{\textbf{Table~\ref{tab:MSSM2to3results}: MSSM $2 \rightarrow 3$ results}}\\
\\
{\bf Process} & ${\mathbf \Delta D}$ \\ \\
 $e_L^- \, W^+ \rightarrow \gamma \, e^- \, x_1^+$ & 6.045$\times 10^ {-7}$\\ 
$u_R \, W^- \rightarrow g \, u_R\, W^-$ & 5.029$\times 10^ {-9}$\\ 
$s\nu_e \, W^- \rightarrow \gamma \, \nu_e \, x_1^-$ & 2.792$\times 10^ {-7}$\\ 
$n_1 \, Z \rightarrow u_R\, \bar u \, Z$ & 8.805$\times 10^ {-10}$\\ 
$u_L \, Z \rightarrow e^+ \, e_L^- \, u$ & 6.262$\times 10^ {-8}$\\ 
$e^+ \, Z \rightarrow e^+ \, e_L^- \, e_R^+$ &$<$ 1.0$\times 10^ {-10}$\\ 
$s\nu_e \, Z \rightarrow e_L^- \, n_1 \, x_1^+$ & 9.288$\times 10^ {-9}$\\ 
$\gamma \, g \rightarrow u_L \, \bar{u}_L \, Z$ & 4.678$\times 10^ {-8}$\\ 
$\gamma \, \bar{u}_L \rightarrow e^+ \, e_R^- \, \bar u$ & 5.273$\times 10^ {-10}$\\ 
$\gamma \, e_L^- \rightarrow e^- \, s\nu_e \, \bar \nu_e$ & 4.001$\times 10^ {-9}$\\ 
$x_1^+ \, x_1^- \rightarrow e^- \, \bar \nu_e \, W^+$ & 2.483$\times 10^ {-7}$\\ 
$e^- \, x_1^+ \rightarrow \nu_e \, W^+ \, x_1^-$ & 3.355$\times 10^ {-7}$\\ 
$s\bar \nu_e \, x_1^+ \rightarrow e^+ \, \nu_e \, \bar \nu_e$ & 6.548$\times 10^ {-8}$\\ 
$e^+ \, x_1^- \rightarrow s\bar \nu_e \, u_R\, \bar{u}_R$ & 1.116$\times 10^ {-10}$\\ 
$n_1 \, n_1 \rightarrow n_1 \, u_R\, \bar u$ & 5.398$\times 10^ {-9}$\\ 
$n_1 \, \bar{u}_L \rightarrow go \, \bar{u}_L \, Z$ & 3.231$\times 10^ {-8}$\\ 
$e_L^- \, n_1 \rightarrow \gamma \, e^-  \, Z$ & 1.842$\times 10^ {-9}$\\ 
$n_1 \, s\nu_e \rightarrow \nu_e \, \nu_e \, \bar \nu_e$ & 2.176$\times 10^ {-9}$\\ 
$g \, u_L \rightarrow e^- \, e_R^+ \, u$ & 3.677$\times 10^ {-10}$\\ 
$go \, go \rightarrow u_R\, \bar{u}_R \, Z$ & 1.089$\times 10^ {-7}$\\ 
$go \, u_R\rightarrow u \, Z \, Z$ & 1.526$\times 10^ {-10}$\\ 
$u \, \bar u \rightarrow \gamma \, u_L \, \bar{u}_L$ & 1.541$\times 10^ {-7}$\\ 
$u \, \bar{u}_R \rightarrow e^+ \, \nu_e \, x_1^-$ & 2.903$\times 10^ {-10}$\\ 
$u \, \bar \nu_e \rightarrow n_1 \, u_L \, \bar \nu_e$ & 9.610$\times 10^ {-10}$\\ 
$u_R\, \bar u \rightarrow \gamma \, s\bar \nu_e \, \nu_e$ & 6.985$\times 10^ {-10}$\\ 
$\bar u \, \nu_e \rightarrow g \, s\nu_e \, \bar{u}_L$ & 1.233$\times 10^ {-8}$\\ 
$e^+ \, u_L \rightarrow e^+ \, u_L \, Z$ & 1.542$\times 10^ {-8}$\\ 
$e_L^+ \, \bar{u}_L \rightarrow s\bar \nu_e \, \bar{u}_L \, W^+$ & 8.574$\times 10^ {-7}$\\ 
$e^+ \, u_R\rightarrow e_R^+ \, u \, Z$ &$<$ 1.0$\times 10^ {-10}$\\ 
$e_R^- \, \bar{u}_R \rightarrow e^- \, g \, \bar u$ & 2.204$\times 10^ {-7}$\\ 
$e^- \, e_L^+ \rightarrow n_1 \, u \, \bar u$ & 2.122$\times 10^ {-8}$\\ 
$e^+ \, e^+ \rightarrow \gamma \, e_L^+ \, e_R^+$ & 1.285$\times 10^ {-9}$\\ 
$e^+ \, s\nu_e \rightarrow \gamma \, n_1 \, W^+$ & 7.022$\times 10^ {-9}$\\ 
$e_R^- \, \nu_e \rightarrow e^- \, e_L^- \, W^+$ & $<$ 1.0$\times 10^ {-10}$\\ 
$\bar \nu_e \, \bar \nu_e \rightarrow e_L^+ \, \bar \nu_e \, x_1^-$ & 4.661$\times 10^ {-8}$\\
\egfbp
\textcolor{white}{\caption{\label{tab:MSSM2to3results}}}
\end{minipage}
\end{table} 
 
\subsection{RS-I processes}

Several models with extra dimensions have recently been proposed. Examples of such theories include the Large Extra Dimensional model (or ADD) \cite{Arkani-Hamed:1998fc,Arkani-Hamed:1999qc} and Randall-Sundrum (RS) models \cite{Randall:1999fk,Randall:1999vf}. These models contain spin-2 particles allowed to propagate through  higher dimensional space-time that couple to the SM particles in a very particular way. To complement the implementation of the ALOHA helicity routines, it is interesting to test and validate also spin-2 routines \cite{Hagiwara:2008zr}. 

To this aim, a simplified version of the Randall-Sundrum type I model is used. Consider a warped 5-dimensional model in which the extra dimension is spatial and compactified in a line segment using the $S_1/Z_2$ symmetry. The propagation on the extra dimension is restricted to spin-2 particles, which will be perceived in 4-dimensions as a tower of excited Kaluza-Klein modes of the same particle with increasing mass. 
We implement a reduced version of the RS-I model: it contains only the first mode of the spin-2 particle explicitly defined, which couples to SM particles.
Validation tests, similar to the ones presented above for the SM and MSSM, are performed. Approximately 500 $2 \rightarrow 2$ and 2000 $2 \rightarrow 3$ processes are tested, and some examples can be found in Tables \ref{tab:RS2to2results} and \ref{tab:RS2to3results}.  Notice the first mode of the spin-2 particle is defined as $y$. The results are again in full agreement. 

\begin{table}
\begin{minipage}[t]{0.5\linewidth}\centering
\bgfbp
\multicolumn{2}{c}{\textbf{Table~\ref{tab:RS2to2results}: RS $2 \rightarrow 2$ results}}\\
\\
{\bf Process} & ${\mathbf   \Delta D}$ \\ \\
$u \, \bar u \rightarrow \gamma \, \gamma$ & $<$ 1.0$\times 10^ {-10}$\\ 
$b~ \, u \rightarrow b~ \, u$ & 1.293$\times 10^ {-8}$\\ 
$u \, \bar \nu_e \rightarrow u \, \bar \nu_e$ & 9.699$\times 10^ {-9}$\\ 
$\bar u \, \bar u \rightarrow \bar u \, \bar u$ & 7.021$\times 10^ {-9}$\\ 
$\gamma \, \bar u \rightarrow \gamma \, \bar u$ & 5.389$\times 10^ {-8}$\\ 
$\bar u \, W^+ \rightarrow \bar u \, W^+$ & 1.836$\times 10^ {-8}$\\ 
$b \, b~ \rightarrow g \, g$ & $<$ 1.0$\times 10^ {-10}$\\ 
$b \, t \rightarrow b \, t$ & 9.656$\times 10^ {-9}$\\ 
$b \, \nu_\mu \rightarrow b \, \nu_\mu$ & 1.293$\times 10^ {-8}$\\ 
$b~ \, t \rightarrow b~ \, t$ & 9.656$\times 10^ {-9}$\\ 
$b~ \, \nu_\mu \rightarrow b~ \, \nu_\mu$ & 1.293$\times 10^ {-8}$\\ 
$t \, t~ \rightarrow u \, \bar u$ & $<$ 1.0$\times 10^ {-10}$\\ 
$t \, t~ \rightarrow \nu_\mu \, \bar \nu_\mu$ & $<$ 1.0$\times 10^ {-10}$\\ 
$t \, \bar \nu_e \rightarrow t \, \bar \nu_e$ & 9.656$\times 10^ {-9}$\\ 
$t~ \, t~ \rightarrow t~ \, t~$ & 7.071$\times 10^ {-9}$\\ 
$t~ \, \bar \nu_\mu \rightarrow t~ \, \bar \nu_\mu$ & 9.656$\times 10^ {-9}$\\ 
$g \, g \rightarrow t \, t~$ & $<$ 1.0$\times 10^ {-10}$\\ 
$g \, g \rightarrow W^+ \, W^-$ & $<$ 1.0$\times 10^ {-10}$\\ 
$g \, Z \rightarrow g \, Z$ & 6.757$\times 10^ {-8}$\\ 
$\gamma \, \gamma \rightarrow g \, g$ & $<$ 1.0$\times 10^ {-10}$\\ 
$\gamma \, \nu_e \rightarrow \gamma \, \nu_e$ & 5.389$\times 10^ {-8}$\\ 
$\gamma \, W^- \rightarrow \gamma \, W^-$ & 6.240$\times 10^ {-8}$\\ 
$\nu_e \, \bar \nu_e \rightarrow \gamma \, \gamma$ & $<$ 1.0$\times 10^ {-10}$\\ 
$\nu_e \, \bar \nu_\mu \rightarrow \nu_e \, \bar \nu_\mu$ & 9.699$\times 10^ {-9}$\\ 
$\bar \nu_e \, \bar \nu_\mu \rightarrow \bar \nu_e \, \bar \nu_\mu$ & 9.699$\times 10^ {-9}$\\ 
$\nu_\mu \, \bar \nu_\mu \rightarrow b \, b~$ & $<$ 1.0$\times 10^ {-10}$\\ 
$\nu_\mu \, \bar \nu_\mu \rightarrow Z \, Z$ & 4.294$\times 10^ {-9}$\\ 
$\bar \nu_\mu \, Z \rightarrow \bar \nu_\mu \, Z$ & 3.671$\times 10^ {-8}$\\ 
$Z \, Z \rightarrow g \, g$ & 1.067$\times 10^ {-8}$\\ 
$W^+ \, Z \rightarrow W^+ \, Z$ & 7.339$\times 10^ {-8}$\\ 
$W^+ \, W^- \rightarrow g \, g$ & 2.677$\times 10^ {-9}$\\ 
$g \, u \rightarrow u \, y$ & $<$ 1.0$\times 10^ {-10}$\\ 
$b \, y \rightarrow b \, g$ & $<$ 1.0$\times 10^ {-10}$\\ 
$\bar{t} \, y \rightarrow g \, \bar{t}$ & $<$ 1.0$\times 10^ {-10}$\\ 
$g \, y \rightarrow t \, \bar{t}$ & $<$ 1.0$\times 10^ {-10}$\\ 
$g \, t \rightarrow t \, y$ & $<$ 1.0$\times 10^ {-10}$\\ 
$g \, g \rightarrow g \, y$ & $<$ 1.0$\times 10^ {-10}$\\  
\egfbp
\textcolor{white}{\caption{\label{tab:RS2to2results}}}
 \end{minipage}
\hspace{0.5cm}
\begin{minipage}[t]{0.5\linewidth}
\centering
\bgfbp
\multicolumn{2}{c}{\textbf{Table~\ref{tab:RS2to3results}: RS $2 \rightarrow 3$ results}}\\
\\
{\bf Process} & ${\mathbf \Delta D}$ \\ \\
$\gamma \, g \rightarrow \gamma \, u \, \bar u$ & 4.431$\times 10^ {-8}$\\ 
$g \, g \rightarrow d \, \bar d \, g$ & 1.412$\times 10^ {-9}$\\ 
$\gamma \, \gamma \rightarrow g \, g \, g$ & 8.610$\times 10^ {-10}$\\ 
$\gamma \, g \rightarrow \gamma \, b \, b~$ & 4.426$\times 10^ {-8}$\\ 
$\gamma \, t \rightarrow \gamma \, g \, t$ & 2.749$\times 10^ {-8}$\\ 
$g \, g \rightarrow g \, g \, g$ & 3.175$\times 10^ {-8}$\\ 
$g \, g \rightarrow e^+ \, e^- \, g$ & 9.659$\times 10^ {-10}$\\ 
$b \, g \rightarrow \gamma \, \gamma \, b$ & 1.449$\times 10^ {-8}$\\ 
$b~ \, g \rightarrow b~ \, e^+ \, e^- $ & 9.753$\times 10^ {-8}$\\ 
$g \, t \rightarrow t \, \nu_e \, \bar \nu_e$ & 4.294$\times 10^ {-9}$\\ 
$g \, t~ \rightarrow g \, g \, t~$ & 2.231$\times 10^ {-8}$\\ 
$g \, \nu_e \rightarrow b \, b~ \, \nu_e$ & 9.876$\times 10^ {-8}$\\ 
$g \, \bar \nu_e \rightarrow t \, t~ \, \bar \nu_e$ & 1.019$\times 10^ {-7}$\\ 
$e^+ \, g \rightarrow e^+ \, g \, g$ & 1.251$\times 10^ {-8}$\\ 
$g \, W^+ \rightarrow g \, g \, W^+$ & 4.333$\times 10^ {-9}$\\ 
$b \, b~ \rightarrow \gamma \, \gamma \, g$ & 6.926$\times 10^ {-9}$\\ 
$b \, b~ \rightarrow g \, \nu_e \, \bar \nu_e$ & 2.382$\times 10^ {-7}$\\ 
$b \, \nu_e \rightarrow b \, g \, \nu_e$ & 3.671$\times 10^ {-8}$\\ 
$b~ \, b~ \rightarrow b~ \, b~ \, g$ & 2.067$\times 10^ {-7}$\\ 
$b~ \, e^+ \rightarrow b~ \, e^+ \, g$ & 3.529$\times 10^ {-9}$\\ 
$t \, t~ \rightarrow b \, b~ \, g$ & 3.815$\times 10^ {-10}$\\ 
$t \, t~ \rightarrow g \, W^+ \, W^-$ & 4.773$\times 10^ {-7}$\\ 
$e^+ \, t \rightarrow e^+ \, g \, t$ & 3.378$\times 10^ {-9}$\\ 
$e^+ \, t~ \rightarrow e^+ \, g \, t~$ & 3.378$\times 10^ {-9}$\\ 
$e^+ \, e^- \rightarrow g \, g \, g$ & 1.903$\times 10^ {-10}$\\ 
$Z \, Z \rightarrow b \, b~ \, g$ & 4.314$\times 10^ {-9}$\\ 
$W^+ \, W^- \rightarrow g \, t \, t~$ & 1.595$\times 10^ {-9}$\\ 
$\bar d \, u \rightarrow \bar d \, g \, u$ & 8.234$\times 10^ {-9}$\\ 
$g \, u \rightarrow d \, \bar d \, u$ & 1.012$\times 10^ {-8}$\\ 
$\bar u \, \bar u \rightarrow g \, \bar u \, \bar u$ & 1.421$\times 10^ {-8}$\\ 
$\gamma \, \bar u \rightarrow \gamma \, g \, \bar u$ & 2.774$\times 10^ {-8}$\\ 
$\bar t \, y \rightarrow g \, g \, \bar t$ & 9.766$\times 10^ {-10}$\\ 
$g \, g \rightarrow g \, g \, y$ & 3.979$\times 10^ {-10}$\\ 
$b \, g \rightarrow b \, g \, y$ & 8.009$\times 10^ {-10}$\\ 
$b \, \bar b \rightarrow b \, \bar b \, y$ & 1.864$\times 10^ {-9}$\\ 
$\bar b \, \bar t \rightarrow \bar b \, \bar t \, y$ & 7.411$\times 10^ {-10}$\\ 
$t \, \bar t \rightarrow b \, \bar b \, y$ & 1.928$\times 10^ {-10}$\\ 
\egfbp
\textcolor{white}{\caption{\label{tab:RS2to3results}}}
\end{minipage}
\end{table} 

\subsection{Beyond the HELAS library}

An important achievement of \aloha~is the ability to create an helicity routine for any type of interaction, even ones not found in the Standard Model. In this context, the \aloha~package, via \madgraph~5, has already been used for some analyses. 

As an example we cite the Strongly Interacting Light Higgs model (SILH). Within this theory, the Higgs is the Goldstone Boson linked to a new strongly interacting sector. This model has been implemented into \ufo~and presented in~\cite{Degrande:2011ua}.  The authors provide a validation example where the partial decay width of the Higgs into two $W$ bosons is computed. This decay is modified by a Non-Standard Model interaction corresponding to the following Lorentz structure:
\begin{verbatim}
VVS8 = Lorentz(name = 'VVS8', 
       spins = [ 3, 3, 1 ], 
       structure = 'P(1,3)*P(2,1) + P(1,2)*P(2,3) 
                   - P(-1,1)*P(-1,3)*Metric(1,2) 
                   - P(-1,2)*P(-1,3)*Metric(1,2)').
\end{verbatim}
This structure corresponds to a dimension 6 operator (and should therefore be proportional to the vacuum expectation value).

The value for the partial decay width is computed by \madgraph~version 5, which uses \aloha, and compared to results previously known. 
This example not only validates the \ufo~model but also the routines automatically generated by \aloha.

\begin{figure}
\centering
\includegraphics[trim=0 300 0 200,width=10cm]{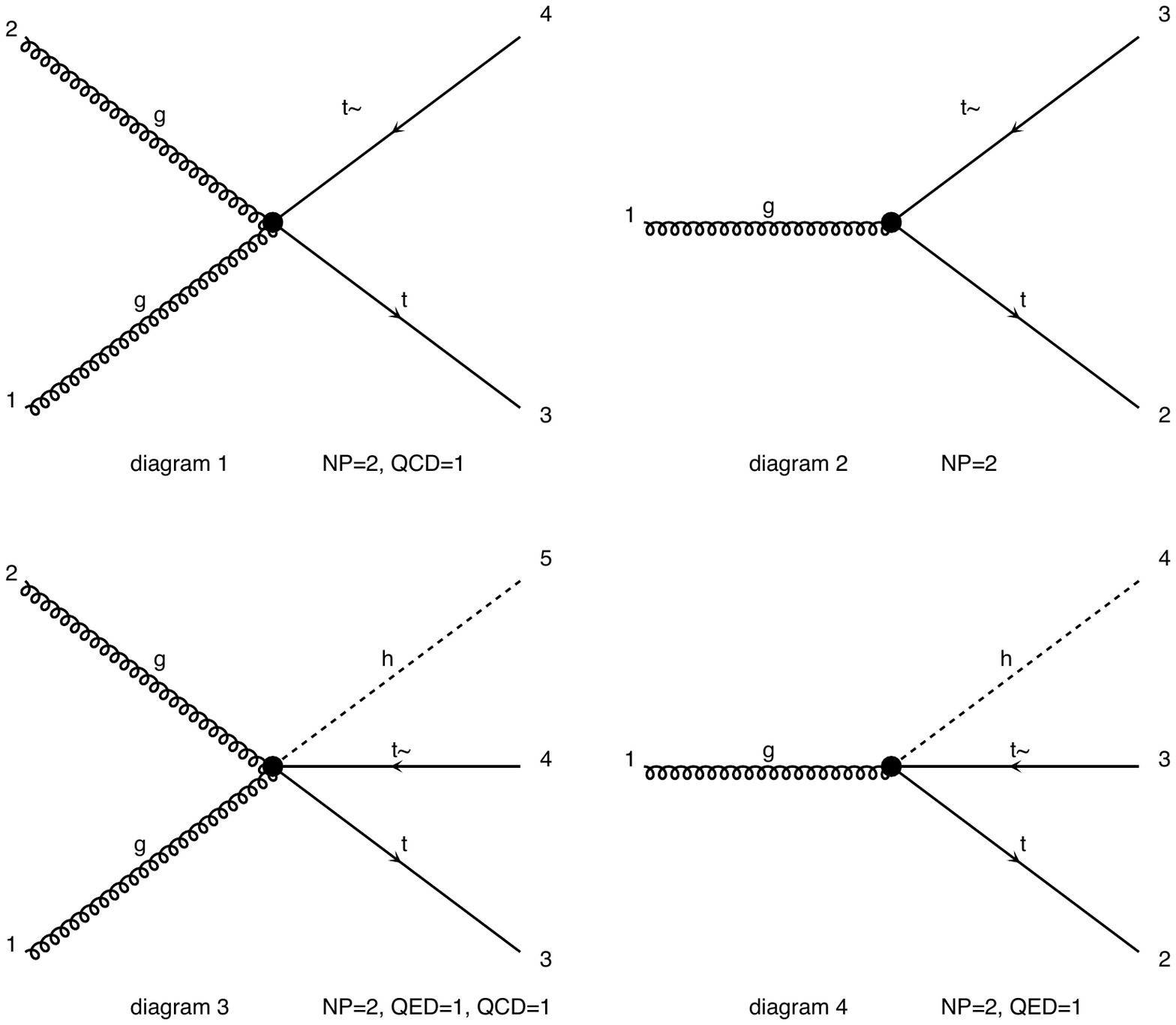}
\caption{Interactions induced by the chromomagnetic operator. }
\label{fig:chromo_diag}
\end{figure}

A second example can be seen in the recent study on effective operators that modify the top production and decay \cite{Degrande:2010kt, Degrande:2011rt}. These analyses are motivated by the precise result on the top sector provided by both the Tevatron and the LHC, and which allow one to constrain the higher dimensional operators. 
For instance, one of these operators is known as the chromo-magnetic operator \cite{Degrande:2010kt}:
\begin{equation}
\mathcal{L} = \frac{(H\bar Q)\sigma^{\mu\nu}T^At G^A_{\mu\nu}}{\Lambda^2} + h.c., \nonumber
\end{equation}
where $\Lambda$ represents the cutoff of the effective theory.
Adding such a term to the Lagrangian of the Standard Model leads to additional interactions shown by Fig.~\ref{fig:chromo_diag}. 
These additional interactions are not covered by the traditional \helas~library. 
In collaboration with the authors of \cite{Degrande:2010kt}, we check the \aloha~results against i) their analytical computation, and ii) their private implementation of the \helas~routines inside the \madgraph~version 4 framework. A very good agreement is again obtained.

The implementation of  \aloha~inside \madgraph~version 5, combined with the ability to automatically generate  the \ufo~model from a general Lagrangian via \feynrules, \aloha~has been used and tested on a wide class of different models. The list of currently available models is presented on the \feynrules~web-page.
These models include, for example, the Higss Effective Theory, the MUED model, the NMSSM and the RMSSM.

\subsection{Speed comparison}

In addition to the validity checks carried out to test the accuracy of the \aloha~routines, one extra comparison is performed to check the speed of the new routines. Table \ref{tab:Speed} shows the time required to compute the squared matrix element using the original optimized \helas~routines and these generated by \aloha. The third column shows the time ratio (\aloha/\helas) revealing that the automatically generated routines are typically within a factor of two in speed of the optimized routines.

\begin{table}
\bgfbz
\multicolumn{4}{c}{\textbf{Table~\ref{tab:Speed}: Computational time (s) to produce values for the  }}\\
\multicolumn{4}{c}{\textbf{$|\mathcal{M}|^ 2$ using either \aloha~or \helas }}\\
\\
Process & ALOHA & HELAS & ALOHA/HELAS \\ \\
$u \, \bar u \rightarrow u \, \bar u \, u \, \bar u$ & 0.2492& 0.1169 & 2.1 \\ 
$u \, \bar u \rightarrow e^- \, e^+ \, g \, g$ & 0.1922& 0.107 & 1.8 \\ 
$e^- \, u \rightarrow e^- \, g \, g \, u$ & 0.1871& 0.1057 & 1.8 \\ 
$g \, u \rightarrow g \, u \, u \, \bar u$ & 0.2286& 0.1105 & 2.1 \\ 
$\bar u \, \bar u \rightarrow e^+ \, e^- \, \bar u \, \bar u$ & 0.2026& 0.1062 & 1.9 \\ 
$g \, g \rightarrow g \, g \, g$ & 0.1966 & 0.1035 & 1.9 \\ \egfbz
\textcolor{white}{\caption{\label{tab:Speed}}}
\end{table} 

\section{Summary and perspectives}
\label{sec:summary}

The automation of cross section calculations and  event generation at leading order (LO), as well as at higher orders, from the Lagrangian level requires several non-trivial steps. In this paper, we have presented a code that automatically generates the helicity routines required by a MC event generator in several widely used formats (\fortran, \cpp~and \python). This is an important step in achieving a fully automatized chain from the Feynman rules, or the Lagrangian level, up to the event generation level.

\aloha~represents a non-trivial extension of the current \helas~library which is limited to renormalizable interactions and fixed Lorentz structures. 
The code is extremely flexible and easily extendible.  A first simple example is the implementation of the complex-mass scheme \cite{Denner:2006ic} for the gauge-invariant computation of amplitudes with unstable intermediate particles (such as top, $W$ and $Z$). As a further example, the current implementation of the Lorentz algebra assumes 4 space-time dimensions, but it could be easily generalized to any other number of even dimensions, as the algebra is symbolic and its actual representation enters only at the final stage of the output writing.

In addition, the implementation of the so-called $R_2$ Feynman rules, which are needed for the automatic loop computation within the Cuttools method \cite{Ossola:2007ax}, can be easily obtained. The necessary modifications for the routines to run on GPU's can be included, and the corresponding {\sc Heget} \cite{Hagiwara:2009aq, Hagiwara:2009cy} library can be automatically generated for any BSM model. Finally, extensions to matrix element generator other than \madgraph, such as that used in \herwig++, is straightforward. Work in these directions is on-going.

\section{Acknowledgments}
The authors would like to thank Johan Alwall and Celine Degrande for their help and support during all phases of this project. We would also like to thank Kaoru Hagiwara, Kentarou Mawatari, Qiang Li and Michel Herquet for their important support.  
A big thanks to the \feynrules, \madgraph~and \ufo~teams without whom this work would not be possible, and to Benjamin Fuks and Claude Duhr for their overall assistance within this project.
Will Link was supported in part by NSF grant PHY-0705682. OM is Chercheur scientifique 
logistique postdoctoral F.R.S-FNRS, Belgium.
This work is supported in part by the FWO - Vlaanderen, Project number  G.0651.11, and in part by the Federal Office for Scientific, Technical and Cultural Affairs through the `Interuniversity Attraction Poles Programme'  Belgian Science Policy P6/11-P and by the IISN MadGraph convention 4.4511.10.

\appendix

\section{Argument convention}
 \label{app:convention}

Since the \fortran, \python~and \cpp~ languages are distinct languages, there are small differences in the procedure used by each language to call the \aloha~functions. 
Table \ref{tab:Argument} presents the different types of arguments, and Table \ref{tab:calling} presents various examples on how the routines should be called in each language and
what is the equivalent analytical formula.

\begin{table}
\bgfbz
\begin{tabular}{l|cl l l}
\multicolumn{5}{c}{\textbf{Table~\ref{tab:Argument}:  Argument format in different output languages}}\\
\multicolumn{5}{c}{}\\
\textbf{argument} & \textbf{symbol}& \textbf{\python} & \textbf{\fortran}                     & \textbf{\cpp} \\
\hline
mass                            &M    & float                    & real       &  complex \\
width                             &W & float                     & real       &  complex \\
coupling                         &Coup& complex              & complex        & complex \\
scalar wavefunctions  &S &   list (3)      &  complex(3)   & complex[3] \\
fermion wavefunctions &F  &    list (6)      &  complex(6)   & complex[6] \\
vector wavefunctions     &V&   list (6)      &  complex(6)   & complex[6] \\
tensor wavefunctions    &T&    list (18)    &  complex(18) & complex[18] \\
Amplitude                      &A&   complex  &  complex         & complex \\ 
\multicolumn{5}{c}{}\\
\multicolumn{5}{c}{\begin{minipage}{0.9\textwidth}
All \fortran, \cpp~arguments are double precision. The arguments in \python~are only informative, since \python~is dynamically typed. 
However, the list of wavefunctions should be initialized with the correct number of elements.\end{minipage}
}\\
\end{tabular}
\egfbz
\textcolor{white}{\caption{\label{tab:Argument}}}

\end{table}

\begin{table}
\bgfbz
\begin{tabular}{l l l}
\multicolumn{3}{c}{ \textbf{Table~\ref{tab:calling}: Syntax Examples} }\\
\\
\textbf{VVSS\_0}\\
&\textbf{analytical:}&  $Coup * V_1^\mu \Gamma_{\mu\nu} V_2^\nu S_3 S_4$\\
&\python& A = VVSS\_0(V1, V2, S3, S4, Coup)\\
&\fortran/ \cpp& VVSS\_0(V1, V2, S3, S4, Coup, A)\\
\\
\textbf{FFV\_0}\\
  & \textbf{analytical:}& $Coup * \bar F_2 \Gamma_{\mu} F_1V_3^\mu$\\
&\python& A = FFV\_0(F1, F2, V3, Coup)\\
 &\fortran/ \cpp& FFV\_0(F1, F2, V3, Coup, A)\\
\\
\textbf{FFV\_1}\\
  & \textbf{analytical:}& $Coup * \bar F_2 \Gamma_{\mu} P^f_1V_3^\mu$\\
&\python& F1 = FFV\_1(F2, V3, Coup, M1,W1)\\
 &\fortran/ \cpp& FFV\_1(F2, V3, Coup, M1, W1, F1)\\
\\
\textbf{FFVC1\_0}\\
  & \textbf{analytical:}& $Coup * \bar F_1 C \Gamma_{\mu}^T C^\dagger F_2 V_3^\mu$\\
&\python& A = FFVC1\_0(F2, F1, V3, Coup)\\
 &\fortran/ \cpp& FFVC1\_0(F2, F1, V3, Coup, A)\\
\\
\textbf{FFVC1\_1}\\
  & \textbf{analytical:}& $Coup * \bar F_1 C \Gamma_{\mu}^T C^\dagger P^f_2 V_3^\mu$\\
&\python& F2 = FFVC1\_0( F1, V3, Coup, M2, W2)\\
 &\fortran/ \cpp& FFVC1\_0(F1, V3, Coup, M2, W2, F2)\\
\\
\textbf{VVV1\_2\_0}\\
  & \textbf{analytical:}& $ V_1^\mu V_2^\nu V_3^\rho ( Coup_1 \Gamma^1_{\mu\nu\rho}   + Coup_2 *   \Gamma^2_{\mu\nu\rho})  $\\
&\python& A = VVV4\_5\_0( V1,V2, V3,Coup1, Coup2)\\
 &\fortran/ \cpp& VVV4\_5\_0(V1,V2, V3, Coup1, Coup2, A)\\
\\
\multicolumn{3}{c}{\begin{minipage}{0.9\textwidth}
$C$ stands for the conjugate matrix, $\Gamma$ stands for the Lorentz structures associated to the vertex, $P^f_i$ is the (fermion) propagator associated to the particle $i$. Other symbols are defined in Table \ref{tab:Argument}.
Note in particular that the order of the arguments in the case of conjugate routines (FFVC1) follow the incoming/outgoing convention (and not the particle order).
The last routine (VVV1\_2\_0) is an example of routines with multiple couplings, corresponding to two Lorentz structures. 
\end{minipage}}
\end{tabular}
\egfbz

\textcolor{white}{\caption{\label{tab:calling}}}
\end{table}

\section{Structure of the computation}
\label{app:analytical}

In order to perform the necessary analytical computations in \python, we design a dedicated analytical package.
The fundamental object of this package is the {\tt Variable} object, which contains three different pieces of information:
\begin{description}
\item[prefactor] The numerical value by which the variable is multiplied. \texttt{[default=1]}
\item[power] The power of the variable. \texttt{[default=1]}
\item[name] A string representing the variable.
\end{description}
For instance, $ 3 x^2$ is  {\tt Variable(name='x', prefactor=3, power=2)}. 
{\tt Variable} objects with identical names are assumed to be referring to the same mathematical variable. Therefore,  addition and multiplication are performed as usual. For example:
\begin{verbatim} 
   Variable(name='x', prefactor =3) + Variable(name='x') 
                            = Variable(name='x', prefactor =4)
\end{verbatim}
 
In the case of a sum (product) between different variables, the sum (product) returns an object which is essentially the list of the summed (multiplied) variables. 
\begin{verbatim}
    Variable(name='x') +  Variable(name='y') 
                        = [Variable(name='x'),  Variable(name='y')]
\end{verbatim}
With such a structure it is possible to represent any type of scalar expression.

Non-scalar objects (like $p^\mu$, $\gamma^\mu$, ...) have basically the same structure, but slightly extended.  There are two additional properties:
 \texttt{lorentz\_ind} and \texttt{spin\_ind}, which are the list of Lorentz and spinor indices linked to the object.  Each object is related to a representation, \ie, each component should be linked to a scalar expression which can be a number, a variable or even an expression.

\aloha~inputs are based on Lorentz objects, and the calculations are performed in seven steps:
\begin{enumerate}
\item Change of momentum convention in order to pass from the \ufo~into the \helas~convention.
\item Multiplication of the Lorentz expression (from \ufo) to the wavefunction or propagator as required.
\item Evaluation of the expression at the abstract level and simplification where appropriated (to adjust prefactors, \etc).
\item Use of the representation of each object in order to create a representation of the result.
\item Simplification of each component.
\item Factorization of each component to have an efficient way to evaluate the result.
\item Writing the routine.
\end{enumerate}
Once the final expression is obtained, the result is passed to a specific routine which writes out the corresponding \helas~routines.  

\aloha~ primarily uses 4 files that correspond to the different steps of the computation:
\begin{description}
\item[create\textunderscore aloha.py:] contains the driver of the code and the part needed to build the  different routines associated with each lorentz structure.
\item[aloha\textunderscore lib.py:] contains all of the code needed for the analytical computations and for the evaluation of the representation of the object.
\item[aloha\textunderscore object.py:]  contains the definition of all basic objects. The current list of basic objects are displayed in Table~\ref{tab:lorentz_structure}. Adding a new object is straightforward since it is enough to specify the index of the objects and the representation. 
\item[aloha\textunderscore writers.py:] contains the code that writes all \aloha~functions in different languages (currently \python, \fortran~and \cpp).
\end{description}


\bibliographystyle{elsarticle-num}
\bibliography{database}
\end{document}